# "Conjectural" links in complex networks

A.A. Snarskii [a,b], D.I. Zorinets [a*], D.V. Lande [a,b]

[a] NTUU " Kyiv Polytechnic Institute " Kyiv, Ukraine
[b] Institute for Information Recording NAS of Ukraine

*Abstract*

This paper introduces the concept of Conjectural Link for Complex Networks, in particular, social networks. Conjectural Link we understand as an implicit link, not available in the network, but supposed to be present, based on the characteristics of its topology. It is possible, for example, when in the formal description of the network some connections are skipped due to errors, deliberately hidden or withdrawn (e.g. in the case of partial destruction of the network).

Introduced a parameter that allows ranking the Conjectural Link. The more this parameter - the more likely that this connection should be present in the network.

This paper presents a method of recovery of partially destroyed Complex Networks using Conjectural Links finding.

Presented two methods of finding the node pairs that are not linked directly to one another, but have a great possibility of Conjectural Link communication among themselves: a method based on the determination of the resistance between two nodes, and method based on the computation of the lengths of routes between two nodes.

Several examples of real networks are reviewed and performed a comparison to know network links prediction methods, not intended to find the missing links in already formed networks.

**Introduction**

Simulation of complex networks (CNs) and the research of their parameters [1, 2, 3, 4] comprise a challenging problem. There are a lot of network parameters. Some of them describe networks in whole, e.g., the average node degree and the clustering coefficient. Some other network characteristics, e.g., the node degree distribution function, provide more detailed information. There are parameters that rank the nodes or links in accordance with a certain algorithm. They are known under the common name "centrality" and include, e.g., the degree centrality, the Katz centrality, the Page Rank, and others.

There are also CN parameters which characterize nodes in pairs. For example, in the problem of missing link prediction [5, 6, 7, 8] (sometimes called "Link Prediction" [5, 6, 7, 8]), such parameters are considered which enable one to estimate the probability for a link to emerge between two nodes when the network grows.

In this work, we are interested in such a characteristic for a pair of nodes, which would evaluate how strongly those nodes are connected with each other. We will assume that the larger is the number of paths connecting two nodes and the shorter are those paths, the stronger the nodes are linked together. The simplest variant of this characteristic is the shortest path length between the nodes. When

Corresponding author at: Kyiv Polytechnic Institute, Kyiv, Ukraine. D.I. Zorinets. deniszorinets@gmail.com

the nodes are linked directly, the path length is equal to unity. It is easy to see that such a simple variant of characteristic as the shortest path length between the nodes is not suitable for the description of, e.g., two nodes that are not linked directly, but have many common neighbors. We believe that the introduced characteristic has to be large both for a pair of directly linked nodes and in the case when the nodes are not linked directly, but have plenty of common links with various lengths.

In real networks, e.g., in social ones, the majority of nodes are known not to be linked directly, with the number of existing links being much less than the possible maximum. Therefore, first of all, we are interested in how strongly the nodes that are not linked directly are linked with each other. The absent direct link will be called a "conjectural" link (CL). Accordingly, all CLs can be ranked using a certain numerical parameter. It is important to notice that CLs with large values of this parameter may really exist, but they were overlooked while describing the network, e.g., the social one. It occurred, besides other reasons, because those links were artificially hidden. An example of networks containing "hidden" links can be the so-called "terrorist networks", i.e. networks containing links between terrorists [9].

As CL parameters, we cam use characteristics introduced when studying the Link Prediction problem [5, 6, 7, 8]. The problem of revealing CLs with large parameter values (the top-most ones in the ranking list) is interesting *per se* and allows one, for instance, to find unexpected (conjectural) links between the literary characters or the members of real social networks.

However, not less interesting and important is the fact that the link ranking enables one to address another problem: the restoration of partially destroyed network, e.g., a social one. Unlike a random CL network, in which the nodes are linked with one another absolutely stochastically, in a social network (the most often, these are scale-free networks), they are linked not randomly, which allows them to be ranked according to that or another numerical parameter, e.g., the number of common nearest neighbors.

The network restoration problem can be formulated as follows: On the basis of information available only for the damaged network, as many of removed links as possible have to be found. The network destruction consists in that some links are removed from the network according to a certain (random or definite) criterion.

## 1. Quantitative description of CLs. Ranking parameters $H$ and $G$

There are plenty of various methods for ranking links between nodes in such a way that this ranking would correspond to our concept of a link between two nodes. One of those methods is the selection of all possible paths between two given nodes and the summation of their weights that correspond to their path lengths.

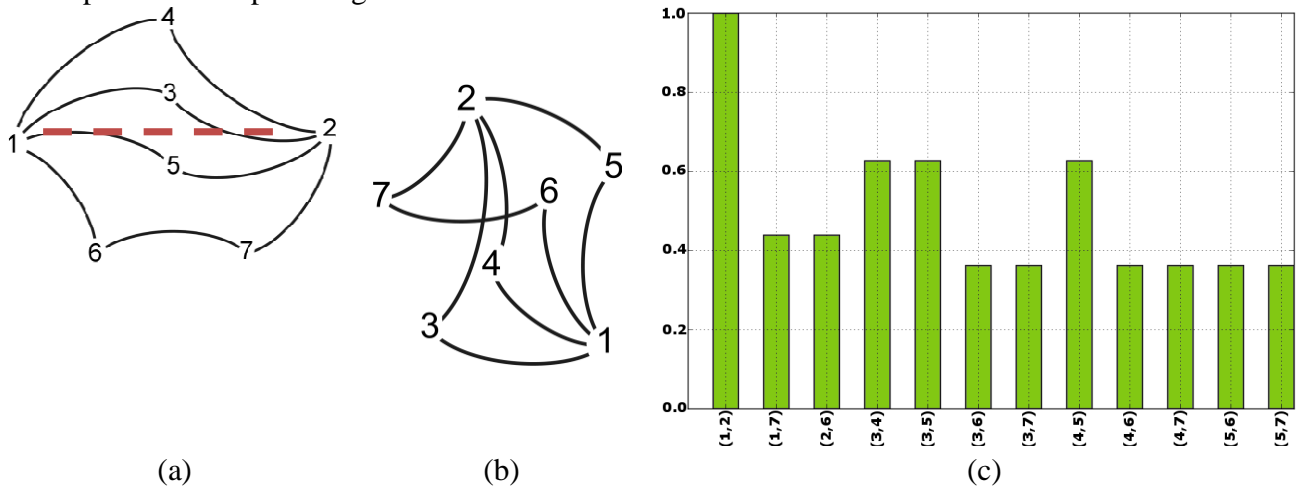

(a) (b) (c)

Fig. 1. A network exhibited in two different forms. (a) Network scheme clearly demonstrating the CL with the largest value of parameter $H$ (the dashed curve). (b) Network scheme for which the observation of CL is rather a complicated task. (c) CL distribution ranked by the parameter $H$.

For illustrative purpose, let us consider a simple network depicted in Fig. 1. Figure 1a makes it is evident that nodes 1 and 2, which are not linked directly, have a lot of common neighbors and, accordingly, a lot of paths between them. A different scheme of the same network, which is shown in Fig. 1b, makes this conclusion, without corresponding calculations, difficult. For large networks, it is impossible to distinguish such links at once. Therefore, calculations using a formal criterion become necessary.

A simple method to evaluate the path number consists in finding the walk number. A walk between the $i$-th and $j$-th nodes is an alternating sequence of links and nodes connecting nodes $i$ and $j$, with the links and nodes in the walk being able to be passed repeatedly [10]. For example, if the $i$-th and $j$-th nodes have a single direct link, the path length between them equals unity, although there may be plenty of other walks between them. If there is a walk equal to one in length, there is also a walk equal to three in length: from $i$ to $j$, backward from $j$ to $i$, and again from $i$ to $j$. The number of walks between the $i$-th and $j$-th nodes with the length $r$ is equal to the $(i,j)$-th element of the adjacency matrix $\mathbf{A}$ raised to the $r$-th power [10].

Let us evaluate the CL ranking parameter as follows:

$$\mathbf{H}_{ij}^{(p)} = \left[\sum_{r=2}^{p} (\alpha \mathbf{A})^r \right]_{ij}, \qquad (1)$$

where the notation $[\ldots]_{ij}$ means the $(i,j)$-th matrix element. Hence, a walk with the length $r$ enters $\mathbf{H}^{(p)}$ with the weight $\alpha$; i.e. the longer is the walk, the smaller contribution it gives to the sum. Note that sum (1) does not include the terms with $r = 0$, because the nodes in the network are not self-connected (a loop), and with $r = 1$, because we do not consider directly linked nodes, i.e. nodes $i$ and $j$ for which $A_{ij} = 0$.

The maximum walk length $p$ can be chosen according to different criteria. In particular, walks with all lengths can be taken into consideration, which equivalent to $p \to \infty$. In this case, summation in Eq. (1) can be carried out to express $\mathbf{H}^{(\infty)}$ in a compact form. For this purpose, let us write $\mathbf{H}^{(\infty)}$ as follows:

$$\mathbf{H}^{(\infty)} = \sum_{r=2}^{\infty} (\alpha \mathbf{A})^r = \sum_{r=0}^{\infty} (\alpha \mathbf{A})^r - (\alpha \mathbf{A})^1 - (\alpha \mathbf{A})^0. \qquad (2)$$

The sum in Eq. (2) can be rewritten:

$$(\mathbf{I} - \alpha \mathbf{A})^{-1} - \mathbf{I} - \alpha \mathbf{A}, \qquad (3)$$

where $\mathbf{I}$ is the unity matrix, and we took into account that $\sum_{r=0}^{\infty}(\alpha \mathbf{A})^r$ is the sum of geometric progression. Now, we should recall that $i \neq j$ for the node pairs, so that $I_{ij}\delta_{ij} = 0$, and that nodes $i$ and $j$ are not connected directly, which means $A_{ij} = 0$ for them. Then, from Eq. (3), we obtain

$$\mathbf{H}^{(\infty)} = (\mathbf{I} - \alpha \mathbf{A})^{-1}, \qquad (4)$$

which, in the case of selected value $\alpha = e^{-2}$, gives

$$\mathbf{H}^{(\infty)} = \left(\mathbf{I} - \frac{1}{e^2}\mathbf{A}\right)^{-1}. \quad (5)$$

Note that $\det(\mathbf{I} - \alpha\mathbf{A}) \neq 0$, so that the inverse matrix $(\mathbf{I} - \alpha\mathbf{A})^{-1}$ can always be calculated. However, one should bear in mind that sum (4) is a series that has to converge in this case. A matrix power series converges if the power series for each of its characteristic values converges [11]. Therefore, Eqs. (4) and (5) can be used only if series (2) converges. This means that the weight $\alpha$ in Eq. (4) cannot be selected arbitrarily. Namely, the product of $\alpha$ and the maximum characteristic value of the adjacency matrix $\mathbf{A}$ must be less than unity.

Note that in [1], an expression similar to Eq. (5) was used while considering the regular equivalence parameter $\boldsymbol{\sigma}$, which obeys the equation

$$\boldsymbol{\sigma} = \alpha\mathbf{A}\boldsymbol{\sigma}\mathbf{A}. \quad (6)$$

If defining the regular equivalence matrix $\boldsymbol{\sigma}$ in such a way [1, 12], its element $\sigma_{ij}$ is supposed to be larger for more similar nodes $i$ and $j$. A high similarity of nodes $i$ and $j$ means that they have neighbors $k$ and $l$ that are similar themselves. It turned out [1] that the quantity $\boldsymbol{\sigma}$ defined by Eq. (6) does not always satisfy this requirement. In this case, to improve the situation, the rightmost matrix $\mathbf{A}$ in the right hand side of Eq. (6) is removed and the unity matrix is introduced as a summand (the both operations are made intentionally), which ultimately brings us to the following equation for $\boldsymbol{\sigma}$:

$$\boldsymbol{\sigma} = \alpha\mathbf{A}\boldsymbol{\sigma} + \mathbf{I}. \quad (7)$$

The solution of this equation,

$$\boldsymbol{\sigma} = \sum_{m=0}^{\infty}(\alpha\mathbf{A})^m = (\mathbf{I} - \alpha\mathbf{A})^{-1}, \quad (8)$$

formally coincides with Eq. (5).

It should be noticed that there is a difference between the regular equivalence $\boldsymbol{\sigma}$ introduced in such a way [Eq. (8)] and the coefficient $\mathbf{H}$. Equation (8) is valid for any pair of nodes $i$ and $j$, including directly connected ones. At the same time, the determination of $[\mathbf{H}^{(p)}]_{ij}$ by Eq. (1) is reduced to Eq. (8) only in the specific case $p \to \infty$. Moreover, it is valid only for indirectly connected nodes.

In Fig. 1c, the ranking distribution of CLs according to the coefficient $[\mathbf{H}^{(p)}]_{ij}$ [Eq. (1)] is shown. The numerical value of $[\mathbf{H}^{(p)}]_{ij}$ does not change already at $p$-values of an order of 10, which means that the contribution of walks longer than 10 practically does not affect sum (1). As is seen from Fig. 1c, the node pairs (1,2) and (3,4) have the largest $[\mathbf{H}^{(p)}]_{ij}$-value, which confirms our qualitative understanding of CLs.

Despite that sometimes the CL ranking in the coefficient $H_{ij}$ and our intuitive feeling about it coincide, the coefficient $H_{ij}$ has a number of shortcomings. In particular, while calculating $H_{ij}$, this is not the lengths of simple paths connecting two nodes that are considered, but the walks between them, in which multiple passages of the same link are allowed. In this case, the longer is the chain, the larger is the number of repeated passages of the same link that are included into the walk. At the same time, the analysis cannot be confined to the shortest paths, because a situation is possible when two nodes are connected by one short path, whereas the other two are not connected by a short path, but have a good many long paths connecting them. Both short and long paths should be taken into account, although longer paths enter the sum with lower weights.

To overcome the indicated shortcomings, we propose to use the conductance $G_{ij}$ between two nodes as an additional ranking coefficient. Let us assume that every link has an electrical resistance equal to unity (accordingly, the conductance of every link also equals unity). Calculating the resistance $R_{ij}$ between nodes $i$ and $j$, we find the conductance by the formula

$$G_{ij} = 1/R_{ij}. \qquad (9)$$

While calculating $R_{ij}$, (i) all possible current paths between nodes $i$ and $j$ are made allowance for; (ii) the current is allowed to run along every link only once; and (iii) it is assumed that the longer is the circuit, the lower current runs along it, and the smaller weight it has in $G_{ij}$. For a simple circuit shown in Fig. 1, the CL ranking according to the parameter **G** also puts node pairs (1,2) and (3,4) on the first place.

Let us proceed now to the consideration of more complicated real networks. Let us consider the social „Karate Club" [13] and the network of characters in the novel "Les Misérables" [14].

The „Karate Club" network [13] is a social network of friend relationships among 34 members of an US university „Karate Club" in 1980s (see Fig. 2).

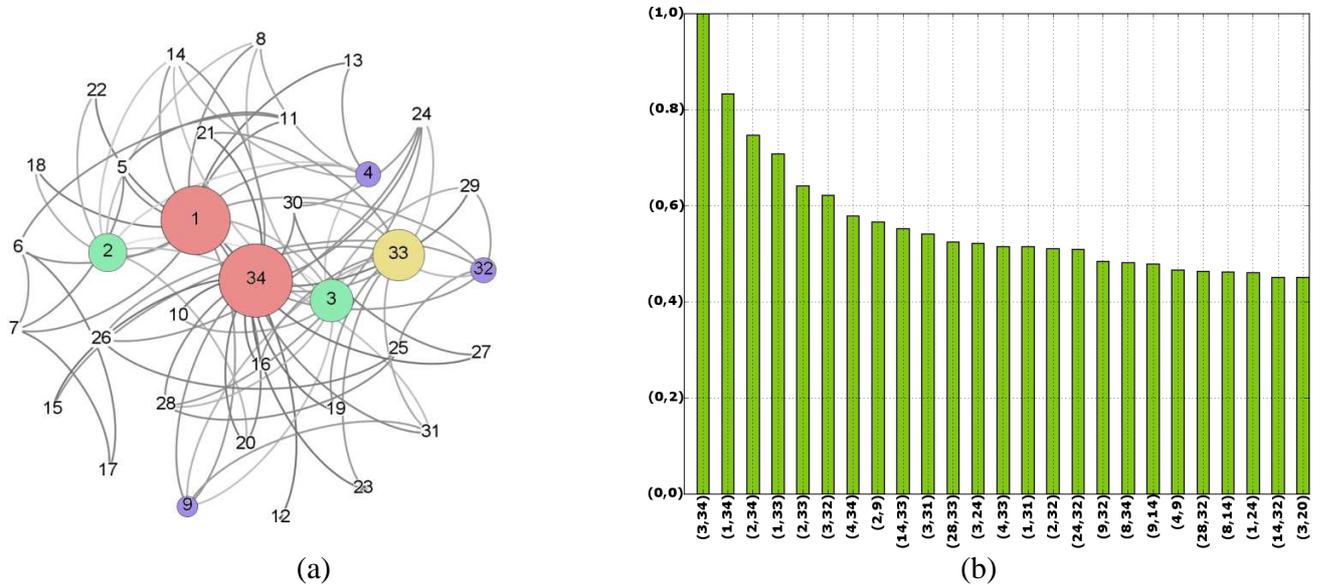

(a) (b)

Fig. 2. The social „Karate Club" network. (a) Its scheme. Bold curves mark the top-rank CLs. Larger circles designate nodes with higher degrees, i.e. with larger numbers of links. (b) CL distribution ranked by the parameter **G**.

As one can see from Fig. 2a, the network of friend relationships among the „Karate Club" members includes such pairs of members, which have many common friends, with the latter not being friends with one another. For instance, let us consider the „Karate Club" members corresponding to nodes 3 and 34. Each of them has rather a high degree. They are not connected directly with each other, but have a lot of indirect links. This fact results in a large $G_{3,34}$-value for those nodes, and their CL is first in the ranked distribution (Fig. 2b). The revealing of similar CL pairs with large values of coefficient **G** can be of interest for sociologists.

Another example is a network of characters in V. Hugo's novel „Les Misérables" [14]. It contains 77 characters and 254 links between them. The total number of possible links equals $77 \times (77-1)/2 = 2926$, i.e. the number of pairs with characters that are not connected directly considerably exceeds the number of pairs with connected characters. When ranking the CLs in **G**, the first place is occupied by the pair Javert--Marius. Hence, it is possible to draw a conclusion that there are certain meaningful relationships among various nodes in the network.

**2. Restoration of a partially destroyed network**

Let we have a network some links in which were removed following a certain rule. Can we restore at least some of destroyed links of the basis of information available only for this damaged network?

It should be marked at once that, if the network has no special structure – for example, if the Erdos-Renyi network [15] is considered, in which links were created randomly (every pair of nodes can be connected with the probability $p$),– its restoration is impossible. There are no considerations about the "deficiency" of links between the nodes in that or another pair. It is also impossible to restore a network, in which a substantial number of links were removed. In this case, information on the network structure is lost. Certainly, by creating links between all pairs of unconnected nodes, it is possible to restore the removed links as well. But the problem is different: to create as few of absent links as possible, with the new links containing as many of removed ones as possible.

Let us consider two scenarios (I and II) of CN restoration. Their initial stages coincide. First, we should prepare a damaged network, i.e. create it. For this purpose: (i) all pairs of connected nodes are ranked by a certain criterion (e.g., by the coefficient **G**), and (ii) some of the first-rank links are removed. Now we can proceed to the network restoration. For this purpose: (iii) all pairs of nodes that are not connected directly are ranked according to a certain criterion, which should not necessarily coincide with that used at the link removal (e.g., by the coefficient **H**), and the links are created between those nodes which obtained the top ranks. In scenario I, links are created until all removed ones become restored. In this case, it is evident that the smaller is the number of created links, the higher is the quality of restoration procedure. As a rule, the number of removed links is unknown. Therefore, another scenario for network restoration is proposed. In scenario II, a definite, initially fixed number of links are created. In this case, the larger number of removed links becomes restored, the higher is the quality of restoration procedure.

Let us consider the parameters of restoration procedure following scenario I. The corresponding coefficient of restoration quality for a network with $N$ nodes can be determined by the formula

$$Q = \frac{\text{number of unconnected node pairs} + \text{the number of removed links} - \text{the of available links}}{\text{the number of available links}} \cdot \frac{\text{the number of removed links}}{\text{the number created links}}$$

$$Q = \frac{M + m - M^+}{M^+} \cdot \frac{m}{m^+}, \qquad (10)$$

Here, $M = N(N-1)/2$ is the number of all possible links in the network, $m$ the number of removed links, $M^+$ the number of available links, and $m^+$ the number of links that have to be created in order to restore all removed links. The nominator of the first fraction in the right hand side of this equation equals to the sum of the number of unconnected node pairs, $M - M^+$, and the number of removed links. The larger is $Q$, the better the network is restored.

For scenario II, let us introduce the following characteristic. After $m$ links have been removed according to a definite algorithm, new $m$ links are created. Let the latter include $K$ links of those which were removed. In this case, the criterion of restoration quality is

$$\eta = m/K \ . \tag{11}$$

Tables A1 and A2 in Appendix A contains $Q$-values evaluated by the **G** and **H** criteria for removed links between the „Karate Club" members and the characters of Hugo's novel "Les Misérables", respectively. Let us consider the coefficients of restoration quality by scenario I for the Scale-free network [16], Hugo's novel "Les Misérables" characters, and the „Karate Club" members. At the first stage, i.e. when ranking the existing links, five ranking methods were selected. Three of them were taken from the Link Prediction problem. Each of those three makes it possible to calculate the coefficient for the link ranking. These are
(i) Jaccard's coefficient [6]

$$J_{ij} = \frac{F(\Gamma(i) \cap \Gamma(j))}{F(\Gamma(i) \cup \Gamma(j))} \ , \tag{12}$$

where $\Gamma(i)$ is the set of nodes connected with the $i$-th node, and $F(\ldots)$ is the number of elements in the set;
(ii) the Adamic-Adar coefficient [6, 7]

$$Ad_{ij} = \sum_{k \in \Gamma(i) \cap \Gamma(j)} \frac{1}{\log F(\Gamma(k))} \ . \tag{13}$$

(iii) the Resourse-Allocation index [17]

$$RA_{ij} = \sum_{k \in \Gamma(i) \cap \Gamma(j)} \frac{1}{F(\Gamma(k))} \ . \tag{14}$$

The other two ranking coefficients are $H_{ij}$ (1) with a final $r$-value and $G_{ij}$ (9). The links were restored using five methods, in which the coefficients $J$, $Ad$, $RA$, $H$, and $G$ were used. While ranking, removing, and creating links, all combinations of five coefficients in pairs were applied.

One of the examined networks was the scale-free one, which was constructed according to the following procedure [15]. A connected network consisting of $m_0$ nodes was created. New nodes were added to the network one by one and became connected to $m \leq m_0$ of existing nodes with the probability $p_i = k_i / (\sum_j k_j)$, where $i$ is the node number, and $k_i$ the degree of the $i$-th node. Then, the links in the obtained realization of the SF network were ranked, e.g., by the $J$-coefficient. First 30 or 10 links were removed, and all non-connected node pairs were ranked, e.g., by the $Ad$-coefficient, and the coefficient of restoration quality, $O_{J,Ad}(10)$, was calculated. The subscript $J, Ad$ specifies the ranking methods used at the link removal and creation (in this case, $J$ and $Ad$, respectively). In total, 25 removal-creation variants were obtained. The number of nodes in the network was $N = 100$. The final realizations of the SF network differed from each other, so that the coefficient $Q$ calculated for each of 30 realizations was averaged.

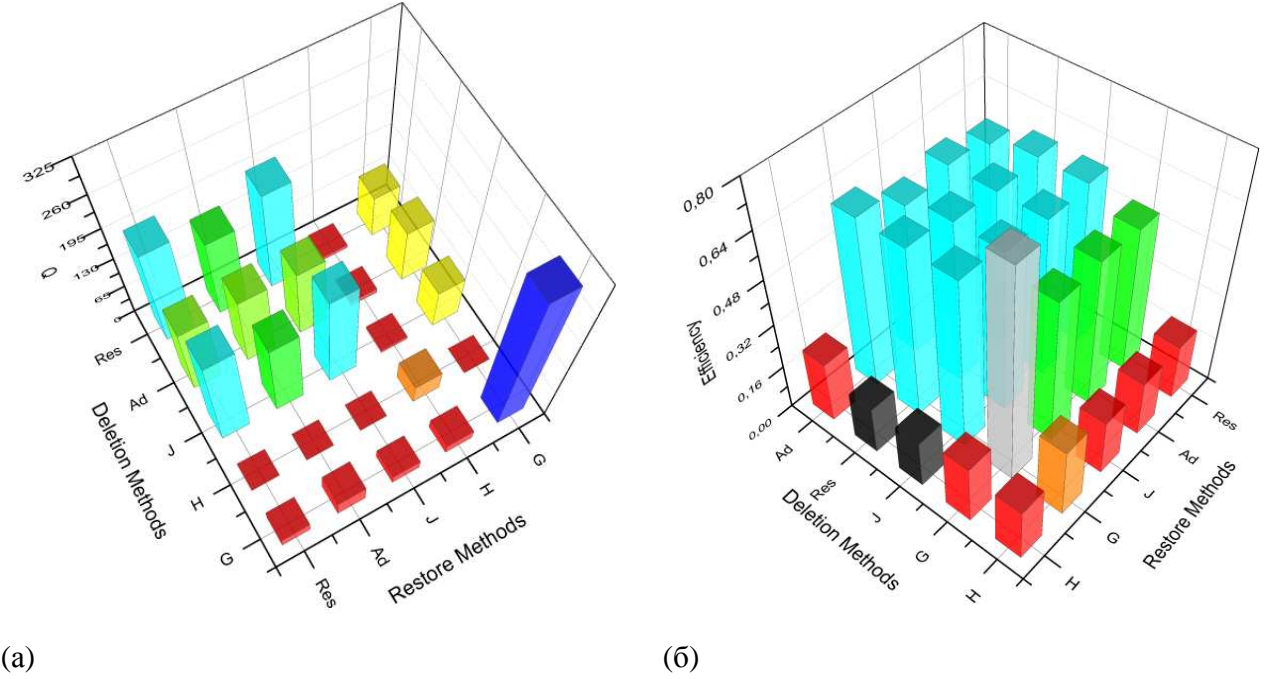

(а) (б)

Fig. 3. The coefficients of network restoration quality $Q$ (a) and $\eta$ (b) for various removal-creation combinations. $G$ is a method based on finding the conductance between two network nodes; $H$ is a method based on finding the set of walks from one node to the other; $J$, $Ad$, and $RA$ are methods based on calculating Jaccard's coefficient, Adamic-Adar coefficient, and Resourse-Allocation index, respectively.

In Fig. 3, the values of all 25 coefficients $Q$'s averaged over the realizations are plotted. From Fig.3, it is evident that the methods $Ad$, $RA$, and $J$ are effective at the restoration of links which were removed by the criteria $Ad$, $RA$, and $J$, but are extremely ineffective at the restoration of links removed with the help of $G$ and $H$ methods. The $Q$-value for the restoration, by the $G$-method, links which were removed by the $G$-method is substantially larger than the others. The $Q$-values corresponding to the removal of links by the $G$ and $H$ methods and their restoration by the $Ad$, $RA$, and $J$ methods are much lower. On the basis of those results, a conclusion can be drawn that the restoration method $G$ is the most effective, and low $Q$- and $\overline{Q}$-values are explained by a substantial dispersion in the position numbers of removed links at their ranking. A conclusion on the network destruction efficiency can also be made. Namely, a network destroyed with the help of $G$-method practically cannot be restored. The $H$-method badly restores removed links, and the links removed by this method are badly restored.

**Conclusions**

To summarize, the so-called conjectural links in complex networks have been studied. Two methods, $H$ and $G$, were proposed for their revealing and ranking. The former is based on finding a set of walks from one node to the other, and the latter on finding the conductance between two network nodes. The problem of network restoration with the help of proposed and some other methods was also considered, and the corresponding results were compared. It was found that in some cases the proposed methods give rise to better results. The quality of network restoration using the $G$-method is higher in comparison with that using the $H$-method, but its result strongly depends on the method used at network destruction.

[1] Newman M. E. J. Networks: an Introduction. Oxford University Press, Computers (2010)
[2] Estrada E. The Sructure of Conplex Netwirks. Oxford University Press (2011)
[3] Freeman L. C. A set of measures of centrality based upon betweenness. Sociometry 40. (1977) 35 41
[4] West D. B. Introduction to Graph Theory. Prentice Hall (2001)
[5] Clauset A. Moore C. Newman M. E. J. Hierarchical structure and the prediction of missing links in networks. Nature 453 (2008) 98 101
[6] Liben-Nowell D. Kleinberg J. The Link-Prediction Problem for Social Networks. Wiley Periodicals (2007) 1019 1031
[7] Lada A. Adamic Eytan Adar. Friends and neighbors on the web. Social Networks, 25(3) (2003) 211 230
[8] Linyuan Lü, Tao Zhou. Link prediction in complex networks: A survey. Phisica A 390 (2011) 1150 1170
[9] Magalingam P. Rao A. Davis S. Identifying a Criminal's Network of Trust. arXiv:1503.04896 [cs.SI] (2015)
[10] Harary F. Graph Theory. Addison-Wesley, Reading, MA (1969)
[11] Korn G. A. Korn T. M. Mathematical Handbook for Scientists and Engineers: Definitions, Theorems, and Formulas for Reference and Review. McGrawHill (1968)
[12] Leicht E. A. Holme P. and Newman M. E. J. Vertex similarity in networks. Phys. Rev E. 73 (2006) 026120
[13] Zachary W. W. An information flow model for conflict and fission in small groups. Journal of Anthropological Research 33 (1977) 452 473
[14] Knuth D. E. The Stanford GraphBase: A Platform for Combinatorial Computing. Addison-Wesley, Reading, MA (1993)
[15] Erdős, P.; Rényi, A. On Random Graphs. I. Publicationes Mathematicae 6 (1959) 290 297
[16] Barabási A.-L. and Albert R. Emergence of scaling in random networks. Science 286 (1999) 509 512
[17] Zhou T. Lu L. Zhang Y.-C. Predicting missing links via local information. Eur. Phys. J. 71 (2009) 623 630
[18] Tizghadam A. Leon-Garcia A. Betweenness Centrality and Resistance Distance in Communication Networks. IEEE Network (2010) 10 16
[19] Noh J. D. Rieger H. Random Walks on Complex Networks. arXiv:cond-mat/0307719 [cond-mat.stat-mech] (2004)
[20] Katz L. A New Status Index Derived from Sociometric. Psychometrika (1953) 39 43
[21] Watts D. J. Strogatz S. H. Collective dynamics of 'small-world' networks. Nature 393 (1998) 440 442
[22] Burt R. S. Structural Holes: The Social Structure of Competition. Harvard University Press (1992)
[23] Dorogovtsev S. N. Mendes J. F. F. Evolution of networks. Advances in Physics 51 (2002) 1079 1187
[24] Snarskii A. A. Lande D. V. Zhenirovskyy M. I. Discovering implicit relations of concepts. Proceedings of XI All-Russian the Conference "Digital Libraries: Advanced Methods and Technologies, Digital Collections" - RCDL'2009 (2009) 46 49.
[25] Blondel V. D. Gajardo A. Heymans M. Senellart P. and Dooren P. V. A measure of similarity between graph vertices: Applications to synonym extraction and web searching. SIAM Review 46 (2004) 647 666

[26] Jeh G. and Widom J. SimRank: A measure of structural-context similarity, in Proceedings of the 8[th] ACP SIGKD International Conference on Knowledge Discovery and Data Mining Association of Computing Machinery (2002) 538 543

[27] Zhenirovskyy M. I. Lande D. V. Snarskii A. A. Detection Implicit Links and G-betweenness. arXiv:1008.4073 [cond-mat.dis-nn](2010)

Appendix A

As an example, let us analyze the restoration of two networks, the network of the „Karate Club" members [12] and the network of Hugo's novel „Les Misérables" characters [13], using the $H$ and $G$ methods. The former method is based on finding a set of walks from one node to the other, and the latter one on finding the conductance between two network nodes.

First, we should prepare each network for the restoration. Namely, we should remove some links to obtain a damaged network. The links are not removed randomly, but only those which have the largest values of the parameter $H$ for the $H$-method and, similarly, the largest $G$-values for the algorithm based on the parameter $G$. For this purpose, we calculate the parameter $H$ (or $G$) for every directly linked node pair and, then, remove those $m$ links from the network which are characterized by the largest $H$- (or $G$-) values.

At the second stage, the parameter $H$ (or $G$) is calculated for all directly non-linked node pairs in the damaged network, and those pairs are ranked once more. Now, we can compare the ranks of node pairs before and after the removal. The corresponding results obtained in the cases $m=10$, 5, and 2 are quoted in Table A1 for the „Karate Club" network and in Table A2 for the „Les Misérables" network. The column "Link" in both tables denotes node pairs (the links), the column "#" indicates their ranks calculated according to the relevant criterion ($H$ or $G$) before the removal, and the columns "10", "5", and "2" demonstrate their ranks in the damaged network after the removal of 10, 5, or 2 links, respectively, using the corresponding method. For instance (see Table 1, method $G$), the link 1--2, being evaluated according to the criterion $G$, had a rank of 2 among the available links in the undamaged network, and it has a rank of 3 among the absent direct links in the damaged one (with removed 10 top-rank links). The corresponding values for the link 2--3 equal 4 and 77.

| $H$ - method ||||| | $G$ - method |||||
|---|---|---|---|---|---|---|---|---|---|---|
| # | Link | 10 | 5 | 2 | | # | Link | 10 | 5 | 2 |
| 1 | 33 - 34 | 1 | 1 | 1 | | 1 | 33 - 34 | 1 | 1 | 1 |
| 2 | 1 - 3 | 24 | 5 | 2 | | 2 | 1 - 2 | 3 | 2 | 3 |
| 3 | 1 - 2 | 3 | 4 | | | 3 | 1 - 3 | 7 | 5 | |
| 4 | 1 - 4 | 17 | 15 | | $Q_{10} = 0.63$ | 4 | 2 - 3 | 77 | 8 | | $Q_{10} = 0.62$ |
| 5 | 2 - 3 | 25 | 16 | | $Q_5 = 1.88$ | 5 | 3 - 33 | 6 | 7 | | $Q_5 = 2.28$ |
| 6 | 1 - 14 | 98 | | | $Q_2 = 6.21$ | 6 | 1 - 4 | 20 | | | $Q_2 = 4.17$ |
| 7 | 9 - 34 | 5 | | | | 7 | 32 - 34 | 5 | | |
| 8 | 3 - 33 | 34 | | | | 8 | 9 - 34 | 9 | | |
| 9 | 3 - 4 | 57 | | | | 9 | 2 - 4 | 28 | | |
| 10 | 32 - 34 | 6 | | | | 10 | 1 - 14 | 19 | | |

Table A1. Data on the restoration of „Karate Club" network.

One can see from Table A1 that the restoration of the „Karate Club" network using the $G$-method is more successful than using the $H$-method. In particular, the $G$-method restored 6 of 10 removed links, whereas the $H$-one only 4. Hence, in the case when 10 links are created, the parameter

$\eta = 0.6$ for the $G$-method and 0.4 for the $H$-one.

The coefficient of restoration quality $Q$ is calculated differently. For instance, in the initial network, the link 2--3 had a rank of 5 evaluated by the criterion $H$. After the removal of $m = 10$ top-rank links, this link has a rank of 25, also evaluated by the method $H$, among the absent links. This means that 25 links have to be created for the latter to include the link 2--3 (to restore it). Among all absent links in the damaged network, we must select the link that has the largest rank. According to Table A1 (the method $H$), the link 1--14 has a rank of 98. Since the rank of this link is the largest, it means that $m^+$, i.e. the number of new links that have to be created to restore all removed ones, equals 98. Then, according to Eq. (10), we obtain $Q_{10} \approx 42.3$. Analogously, for the method $G$, after the removal of 10 links, the link 2--3 has the maximum rank among them, $m^+ = 77$. Again, according to Eq. (10), we obtain $Q_{10} \approx 0.62$. Hence, the $G$-method turns out much more effective.

| $H$ - method | | | | | | $G$ - method | | | | | |
|---|---|---|---|---|---|---|---|---|---|---|---|
| # | Link | 10 | 5 | 2 | | # | Link | 10 | 5 | 2 | |
| 1 | Gavrosh - Enjorlas | 15 | 1 | 5 | | 1 | Valjan - Javert | 1 | 1 | 1 | |
| 2 | Gavrosh - Marius | 7 | 2 | 19 | | 2 | Gavroch - Marius | 5 | 3 | 2 | |
| 3 | Valjan - Gavrosh | 1 | 3 | | | 3 | Gavroch - Enjorlas | 4 | 2 | | |
| 4 | Gavrosh - Bossuet | 9 | 4 | | | 4 | Valjan - Marius | 2 | 5 | | |
| 5 | Gavrosh - Courfeyrac | 6 | 8 | | $Q_{10} = 42.3$ $Q_5 = 6.32$ $Q_2 = 1,1$ | 5 | Valjan - Thenardier | 3 | 4 | | $Q_{10} = 105.8$ $Q_5 = 10.54$ $Q_2 = 10.53$ |
| 6 | Gavrosh - Joly | 14 | | | | 6 | Valjan - Gavroch | 7 | | | |
| 7 | Gavrosh - Bahorel | 13 | | | | 7 | Marius - Enjorlas | 6 | | | |
| 8 | Marius - Enjorlas | 12 | | | | 8 | Gavroch - Bossuet | 9 | | | |
| 9 | Enjorlas - Bossuet | 4 | | | | 9 | Gavroch - Courfeyrac | 8 | | | |
| 10 | Valjan - Enjorlas | 24 | | | | 10 | Thenardier - Javert | 10 | | | |

Table A2 contains the same information for the „Les Misérables" character network. One can see that its restoration using the $H$-method is rather bad, which is evident from a large dispersion of node pair positions even when only two links were removed. At the same time, the $G$-method always restores all removed links.

Table A2. Data on the restoration of „Les Misérables" character network.